\documentclass[a4paper,11pt,twoside]{article}
\usepackage[INRIA]{lipinria}
\usepackage[latin1]{inputenc}
\usepackage{a4wide}
\usepackage{amsfonts,amsmath,amssymb,amsthm,amstext,latexsym,paralist}
\usepackage{url}
\usepackage{xspace}
\usepackage{algorithm2e}
\usepackage{subfigure}

\usepackage{figlatex}
\newtheorem{lemma}{Lemma}
\newtheorem{theorem}{Theorem}

\newtheorem{definition}{Definition}

\AtBeginDocument{}

\newcommand{\comm}{\textsf{$\delta$}} % comm for stage
\newcommand{\calc}{\textsf{w}}    % calc for  stage
\newcommand{\speed}{\textsf{s}}   % speed of proc
\newcommand{\bw}{\textsf{b}}      % bandwidth of link

\newcommand{\alloc}{\textsf{alloc}}
\newcommand{\period}{T_{\textsf{period}}}
\newcommand{\latency}{T_{\textsf{latency}}}

\newcommand{\COMHOM}{\textit{Communication Homogeneous}\xspace}

\newcommand{\HETPART}{\textsc{Hetero-1D-Partition}\xspace}
\newcommand{\HETPARTDEC}{\textsc{Hetero-1D-Partition-Dec}\xspace}

\newcommand{\n}{n} %\textsf{n}} % number of stages
\newcommand{\p}{p} %\textsf{p}} % number of procs

\newcommand{\link}{\textsf{link}} % number of procs
\renewcommand{\S}{\mathcal{S}} % stage
\newcommand{\II}{\mathcal{I}} % for the proofs

\usepackage{RR}
\begin{document}

\RRInumber{2007-32}
\RRNo{6232}
\RRItitle{Multi-criteria scheduling of pipeline workflows}
\RRItitre{Ordonnancement multi-crit\`ere des workflows pipelin\'es}
\RRIdate{June 2007}
\RRIauthor{Anne Benoit\and Veronika Rehn-Sonigo\and Yves Robert}

\RRIthead{Multi-criteria scheduling of pipeline workflows}
\RRIahead{A. Benoit\and V. Rehn-Sonigo\and Y. Robert}

\RRIkeywords{algorithmic skeletons, pipeline, multi-criteria 
optimization, complexity results, heuristics, heterogeneous 
platforms.}
\RRImotscles{squelettes algorithmiques, pipeline, optimisation
multi-crit\`ere, plates-formes h\'et\'erog\`enes. }

\RRIabstract{Mapping workflow applications onto parallel platforms is a challenging
problem, even for simple application patterns such as pipeline graphs.
Several antagonist criteria should be optimized, such as throughput and
latency (or a combination). In this paper, we study the complexity of
the bi-criteria mapping problem for pipeline graphs on communication
homogeneous platforms.
In particular, we assess the complexity of the well-known
chains-to-chains problem for different-speed processors, which turns
out to be NP-hard.
We provide several efficient polynomial bi-criteria heuristics, and
their relative performance is evaluated through
extensive simulations.}
\RRIresume{L'ordonnancement et l'allocation de workflows sur plates-formes
parall\`eles est un probl\`eme crucial, m\^eme pour des
applications simples comme des graphes en pipeline.
Plusieurs crit\`eres contradictoires doivent \^etre
optimis\'es, tels que le d\'ebit et la latence (ou une combinaison
des deux). 
Dans ce rapport, nous \'etudions la complexit\'e du probl\`eme
de l'ordonnancement bi-crit\`ere pour les graphes en pipeline sur des plates-formes
avec communications homog\`enes.
En particulier nous \'evaluons la complexit\'e du probl\`eme bien
connu  ``chains-on-chains''  pour les processeurs h\'et\'erog\`enes,
un probl\`eme qui s'av\`ere NP-difficile.
Nous proposons plusieurs heuristiques bi-crit\`eres efficaces
en temps polynomial.  Leur performance relative est evalu\'ee
par des simulations intensives. }

\RRItheme{\THNum}
\RRIprojet{GRAAL}
\RRNo{6232}

\RRImaketitle

\section{Introduction}
\label{sec.intro}

Mapping applications onto parallel platforms is a difficult
challenge. Several scheduling
and load-balancing techniques have been developed for homogeneous architectures
(see~\cite{ieee-sched} for a survey) but the advent of heterogeneous
clusters has rendered
the mapping problem even more difficult. Typically, such clusters are
composed of
different-speed processors interconnected either by plain Ethernet
(the low-end version)
or by a high-speed switch (the high-end counterpart), and they
constitute the experimental platform of choice in most academic or
industry research departments.

In this context of heterogeneous platforms, a structured programming
approach rules out many of the problems which the
low-level parallel application developer is usually confronted to,
such as deadlocks or process starvation.
Moreover, many real applications draw from a range of well-known
solution paradigms, such as pipelined or farmed
computations. High-level approaches based on algorithmic
skeletons~\cite{Co02,Ra02} identify such patterns and seek to make it
easy for an application developer to tailor such a paradigm to a
specific problem. A library of skeletons is provided to the programmer,
who can rely on these already coded patterns to express the
communication scheme within its own application.
Moreover, the use of a particular skeleton carries with it considerable
information about implied scheduling dependencies, which we believe
can help address the complex problem of mapping a distributed
application onto a heterogeneous platform.

In this paper, we therefore consider %regular
applications that can be expressed
as algorithmic skeletons, and we focus on the pipeline skeleton, which
is one of the most widely used.
In such workflow applications, a series of data sets (tasks) enter the
input stage and progress from stage to stage until the final result is
computed.
Each stage has its own communication and computation requirements: it
reads an input file
from the previous stage, processes the data and outputs a result to the next stage.
For each data set, initial data is input to the first stage, and final
results are output
from the last stage. The pipeline workflow operates in synchronous
mode: after some latency due to the initialization delay, a new task
is completed every period. The period is defined
as the longest cycle-time to operate a stage. %, and is the inverse of
%the throughput that can be achieved.

Key metrics
for a given workflow are the throughput and the latency. The
throughput measures the aggregate rate of
processing of data, and it is the rate at which data sets can enter the system.
Equivalently, the inverse of the throughput, defined as the period, is
the time interval
required between the beginning of the execution of two consecutive data sets.
The latency is the time elapsed between the beginning and the end of the execution
of a given data set, hence it measures the response time of the system
to process the data set
entirely. Note that it may well be the case
that different data sets have different latencies (because they are
mapped onto different processor sets),
hence the  latency is defined as the maximum
response time over all data sets. Minimizing the latency is
antagonistic to minimizing the period,
and tradeoffs should be found between these criteria.
%Efficient mappings aim at
%the minimization of a single criterion, either the period or the
%latency.
In this paper, we focus on bi-criteria approaches, i.e.
 minimizing the latency under period constraints, or the converse.

The problem of mapping pipeline skeletons onto parallel platforms has
received some attention,
and we survey related work in Section~\ref{sec.related}. In this
paper, we target heterogeneous
clusters, and aim at deriving optimal mappings for a bi-criteria
objective function, i.e. mappings which minimize the
period for a fixed maximum latency, or which minimize the latency for a fixed
maximum period.
Each pipeline stage can be seen as a sequential procedure which may
perform disc accesses or write data in the memory for each task. This
data may be reused from one task to another, and thus the rule of
the game is always to process the tasks in a sequential order within a
stage. Moreover, due to the possible local memory accesses, a given
stage must be mapped onto a single processor: we cannot process half
of the tasks on a processor and the remaining tasks on another without
exchanging intra-stage information, which might be costly and
difficult to implement.
%
%In this paper, we focus on pipeline skeletons and thus
%we enforce the rule that a given stage is mapped onto a single processor.
In other words, a processor
that is assigned a stage will execute the operations required by this
stage (input, computation and output) for all the tasks fed into the
pipeline.

The optimization problem can be stated informally as
follows: which stage to assign to which processor? We require the
mapping to be interval-based, i.e. a processor is assigned an interval
of consecutive stages.
We target \COMHOM platforms, with identical links but different speed
processors, which introduce a first degree of
heterogeneity. Such platforms correspond to networks of workstations
interconnected by a LAN, which constitute the typical experimental
platforms in most academic or research departments.

The main objective of this paper is to assess the complexity of the
bi-criteria mapping problem onto \COMHOM platforms. %We establish the
%complexity result for minimizing the latency (polynomial) and
%minimizing the period (NP-hard),
%
An interesting consequence of one of the new complexity results % proved
%in this paper (see Section~\ref{sec.complexity})
is the following. Given an array of $\n$ elements $a_1, a_2, \ldots,
a_\n$, the well-known chains-to-chains
problem is to partition the array into $p$ intervals whose element
sums are well balanced (technically,
the aim is to minimize the largest sum of the elements of any interval).
This problem has been extensively studied in the literature (see the pioneering papers~\cite{Bokhari88,HansenLih92,olstad95efficient} and the survey~\cite{PinarAykanat2004}).
It amounts to load-balance
$\n$ computations whose ordering must be preserved (hence the restriction to intervals)
onto $p$ identical processors. The advent of heterogeneous clusters
naturally leads to the following generalization:
can we partition the $\n$ elements into $p$ intervals
whose element sums match $p$ prescribed values (the processor speeds)
as closely as possible?
The NP-hardness of this important extension of the chains-to-chains
problem is established in Section~\ref{sec.complexity}.
Thus the bi-criteria mapping problem is NP-hard,
and we derive efficient polynomial bi-criteria heuristics, which
are compared through simulation.

The rest of the paper is organized as follows.
Section~\ref{sec.model} is devoted the presentation of the
target optimization problems. Next in Section~\ref{sec.complexity} we
proceed to the complexity results. In
Section~\ref{sec.heuristics} we introduce several polynomial
heuristics to solve the mapping problem. These heuristics are compared
through simulations, whose
results are analyzed in Section~\ref{sec.experiments}.
Section~\ref{sec.related} is devoted to an
overview of related work. Finally, we state some concluding remarks
in Section~\ref{sec.conclusion}.

\section{Framework}
\label{sec.model}

%We outline in this section the characteristics of the applicative
%framework, as well as the model for the target platform. Next we
%detail the bi-criteria optimization problem.

%\begin{figure}[tbh]
%\begin{center}
%\includegraphics[height=5cm]{fig/clique.fig}
%\end{center}
%\caption{The target platform.} \label{fig.clique}
%\end{figure}

\begin{figure}[tbh]
\begin{center}
\includegraphics{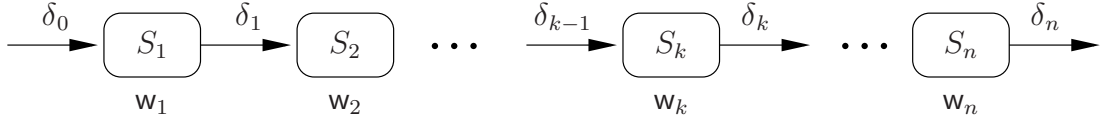}
\end{center}
\caption{The application pipeline.} \label{fig.pipeline}
\end{figure}

{\bf Applicative framework.}
We consider a pipeline of $\n$ stages $\S_k$, $1 \leq k \leq \n$, as illustrated on
Figure~\ref{fig.pipeline}. Tasks are fed into the pipeline and processed from stage to
stage, until they exit the pipeline after the last stage.
The $k$-th stage $\S_k$ receives an input from the previous stage, of size $\comm_{k-1}$,
performs a number of $\calc_k$ computations, and outputs data of size $\comm_{k}$ to the
next stage. The first stage $\S_1$ receives an input of size $\comm_0$ from the outside
world, while the last stage $\S_{\n}$ returns the result, of size $\comm_{\n}$, to the
outside world.

%\subsection{Target platform}

\smallskip
\noindent {\bf Target platform.}
We target a platform %(see Figure~\ref{fig.clique}),
with $\p$ processors
$P_u$, $1 \leq u \leq \p$, fully interconnected as a (virtual) clique. There is a
bidirectional link $\link_{u,v}: P_u \to P_v$ between any processor pair $P_u$ and $P_v$,
of bandwidth $\bw_{u,v}$. Note that we do not need to have a physical link between any
processor pair. Instead, we may have a switch, or even a path composed of several
physical links, to interconnect $P_u$ and $P_v$; in the latter case we would retain the
bandwidth of the slowest link in the path for the value of $\bw_{u,v}$.
In the most general case, we have fully heterogeneous platforms, with different
processor speeds and link capacities, but we restrict in this paper
to \COMHOM platforms with different-speed processors
($\speed_u \neq \speed_v$)
 interconnected by links of same capacities ($\bw_{u,v} = \bw$). They correspond to
 networks of different-speed processors or workstations
 interconnected by either plain Ethernet or by a high-speed
switch, and they  constitute the typical experimental platforms in most academic or industry
research departments.

The speed of processor $P_u$ is denoted as
$\speed_u$, and it takes $X/\speed_u$ time-units for $P_u$ to execute $X$ floating point
operations. We also enforce a linear cost model for communications, hence it takes
$X/\bw$ time-units to send (resp. receive) a message of size $X$ to (resp. from)
$P_v$.
Communications contention is taken care of by enforcing the \emph{one-port}
model~\cite{Bhat99efficient,BhatRagra03}. In this model, a given processor can be
involved in a single communication at any time-step, either a send or a receive. However,
independent communications between distinct processor pairs can take place
simultaneously. The one-port model seems to fit the performance of some current MPI
implementations, which serialize asynchronous MPI sends as soon as message sizes exceed a
few megabytes~\cite{SaifPa2004}.

%\subsection{Mapping problem}
\smallskip
\noindent {\bf Bi-criteria mapping problem.} The general mapping problem
consists in assigning application stages
to platform processors. For the sake of simplicity, we can assume that
each stage $\S_k$ of the application
pipeline is mapped onto a distinct processor (which is possible only if
$\n \leq \p$).
However, such one-to-one mappings may be unduly restrictive, and a
natural extension is to search
for interval mappings, i.e. allocation functions where each participating processor is
assigned an interval of consecutive stages. Intuitively, assigning several consecutive
tasks to the same processors will increase their computational load, but may well
dramatically decrease communication requirements. In fact, the best interval mapping may
turn out to be a one-to-one mapping, or instead may enroll only a very small number of
fast computing processors interconnected by high-speed links.

Interval mappings constitute a natural and useful generalization of
one-to-one mappings (not to speak of situations where $\p < \n$, where
interval mappings are mandatory), and such mappings have been studied
by Subhlock et al.~\cite{subhlock-ppopp95,subhlock-spaa96}.
%
%parler de one-to-one et interval=generalization
%
%It is natural to map
%intervals of consecutive stages onto processors~\cite{subhlock-ppopp95,subhlock-spaa96}.
%Intuitively, assigning several consecutive
%tasks to the same processor will increase their computational load, but may well
%dramatically decrease communication requirements.
The cost model associated to
interval mappings is the following. We search for a partition of $[1..\n]$ into $m \leq \p$
intervals $I_j = [d_j, e_j]$ such that $d_j \leq e_j$ for $1 \leq j \leq m$, $d_1 = 1$,
$d_{j+1}= e_j + 1$ for $1 \leq j \leq m-1$ and $e_m = \n$. Interval $I_j$ is mapped onto
processor $P_{\alloc(j)}$, and the period is expressed as

\begin{equation}
\label{eq.interval.period}
\period = \max_{1 \leq j \leq m} \left\{  \frac{\comm_{d_j -
1}}{\bw}%_{\alloc(j-1),\alloc(j)}}
+ \frac{\sum_{i=d_j}^{e_j}
\calc_{i}}{\speed_{\alloc(j)}} +
\frac{\comm_{e_j}}{\bw}%_{\alloc(j),\alloc(j+1)}}
\right\}
\end{equation}

%\noindent %Here, we assume that $\alloc(0) = \inn$ and $\alloc(m+1) = \out$.
The latency
is obtained by the following expression (data sets traverse all
stages, and only interprocessor communications need be paid for):

\begin{equation}
\label{eq.interval.latency}
\latency = \sum_{1 \leq j \leq m} \left\{  \frac{\comm_{d_j -
1}}{\bw}%_{\alloc(j-1),\alloc(j)}}
+ \frac{\sum_{i=d_j}^{e_j}
\calc_{i}}{\speed_{\alloc(j)}} \right\} +
\frac{\comm_{n}}{\bw}%_{\alloc(j),\alloc(j+1)}}
\end{equation}

The
optimization problem is to determine the best mapping, over all possible
partitions into intervals, and over all processor assignments. The objective can
be to minimize either the period, or the latency, or a combination: given a threshold
period, what is the minimum latency that can be achieved? and the counterpart:
given a threshold latency, what is the minimum period that can be
achieved?

%In this paper, we concentrate on interval-based mappings,
%because they realize the best trade-off between efficiency and
%simplicity. General mappings would require important changes in the model
%and its implementation, and may be too complex to deal with
%efficiently for the application programmer.

\section{Complexity results}
\label{sec.complexity}

To the best of our knowledge, this work is the first to study the
complexity of the bi-criteria optimization problem for an
interval-based mapping of pipeline applications onto \COMHOM platforms.

Minimizing the latency is trivial, while minimizing the period is
NP-hard. Quite interestingly, this last result is a consequence of the
fact that the natural extension of the
chains-to-chains problem~\cite{PinarAykanat2004} to different-speed
processors is NP-hard.

\begin{lemma} The optimal pipeline mapping which minimizes the latency
can be determined in polynomial time.
\end{lemma}

\begin{proof}
The minimum latency can be achieved by mapping the whole interval onto
the fastest processor~$j$, resulting in the latency
 $\left( \sum_{i=1}^n \calc_i \right) / \speed_j$.
If a slower processor is involved in the mapping, the latency
increases, following equation~(\ref{eq.interval.latency}), since part
of the computations will take longer, and communications may occur.
\end{proof}

Thus, minimizing the latency can be done in polynomial time. 
However, it is
not so easy to minimize the period, and we study the heterogeneous 1D
partitioning problem in order to assess the complexity of the period
minimization problem. %First we introduce this new problem.

%Theorem~\ref{th.npc2c} is a consequence of Theorem~\ref{th.hetpart}
%which assesses the complexity of
%the heterogeneous 1D partitioning problem. We introduce this problem
%and prove Theorem~\ref{th.hetpart}
%before returning to the proof of Theorem~\ref{th.npc2c}.

\medskip
Given an array of $\n$ elements $a_1, a_2, \ldots, a_\n$, the 1D partitioning problem,
also known as the chains-to-chains problem,  is to partition the array into $p$ intervals whose element sums are almost identical. More precisely, we search for a partition of $[1..\n]$ into $p$ consecutive intervals
$\II_1, \II_2, \ldots, \II_{\p}$ , where $\II_k = [d_k, e_k]$ and $d_k \leq e_k$ for $1 \leq k \leq p$, $d_1 = 1$,
$d_{k+1}= e_k + 1$ for $1 \leq k \leq p-1$ and $e_p = \n$.
The objective is to minimize
$$\max_{1 \leq k \leq p} \sum_{i \in \II_k} a_i = \max_{1 \leq k \leq p} \sum_{i=d_k}^{e_k} a_i.$$

This problem has been extensively studied in the literature
because it has various applications. In particular, it amounts to load-balance
$\n$ computations whose ordering must be preserved (hence the restriction to intervals)
onto $p$ identical processors. Then each $a_i$ corresponds to the execution time
of the $i$-th task, and the sum of the elements in interval $\II_k$ is the load of the
processor which $\II_k$ is assigned to. Several algorithms and heuristics have been
proposed to solve this load-balancing problem,
including~\cite{Bokhari88,Iqbal91,HansenLih92,IqbalBok95,olstad95efficient}.
We refer the reader to the survey paper by Pinar and Aykanat~\cite{PinarAykanat2004}
for a detailed overview and comparison of the literature.

%For this reason, the problem is also known
%as the chain-to-chain problem, because me map a chain of computations onto
%a chain of resources.

The advent of heterogeneous clusters leads to the following generalization of
the 1D partitioning problem: the goal is to partition the $\n$ elements into $p$ intervals
whose element sums match $p$ prescribed values (the processor speeds) as closely as possible.
Let $\speed_1, \speed_2, \ldots, \speed_{\p}$ denote these values. We search for a partition of $[1..\n]$ into $p$ intervals $\II_k = [d_k, e_k]$ and for a permutation $\sigma$ of $\{1, 2, \ldots, \p \}$,
with the objective to minimize:
$$\max_{1 \leq k \leq p} \frac{\sum_{i \in \II_k} a_i}{\speed_{\sigma(k)}}.$$

Another way to express the problem is that intervals are now weighted by the $\speed_i$ values,
while we had $\speed_i = 1$ for the homogeneous version.
Can we extend the efficient algorithms described in~\cite{PinarAykanat2004}
 to solve the heterogeneous 1D partitioning problem, \HETPART for
short? In fact, the problem seems combinatorial, because of the search
over all possible permutations to weight the intervals. Indeed, we prove the
NP-completeness of (the decision problem associated to) \HETPART.

\begin{definition}[\HETPARTDEC] Given $\n$ elements $a_1, a_2, \ldots, a_{\n}$,
$\p$ values $\speed_1, \speed_2, \ldots, \speed_{\p}$ and a bound $K$, can we find a
partition of $[1..\n]$ into $p$ intervals $\II_1, \II_2, \ldots, \II_{\p}$, with $\II_k = [d_k, e_k]$ and $d_k \leq e_k$ for $1 \leq k \leq p$, $d_1 = 1$,
$d_{k+1}= e_k + 1$ for $1 \leq k \leq p-1$ and $e_p = \n$,  and a
permutation $\sigma$ of $\{1, 2, \ldots, \p \}$,
such that
$$\max_{1 \leq k \leq p} \frac{\sum_{i \in \II_k} a_i}{\speed_{\sigma(k)}} \leq K \quad ?$$
\end{definition}

\begin{theorem}
\label{th.hetpart}
The \HETPARTDEC problem is NP-complete.
\end{theorem}

\begin{proof}

%{\bf (Theorem 1). }
The \HETPARTDEC problem clearly belongs to the class NP: given a solution, it is
easy to verify in polynomial time that the partition into $\p$ intervals is valid
and that the maximum sum of the elements in a given interval divided by the corresponding
$\speed$ value does not exceed the bound $K$.
To establish the
completeness, we use a reduction from NUMERICAL MATCHING WITH TARGET
SUMS (\textsc{NMWTS}), which is NP-complete
in the strong sense~\cite{GareyJohnson}. We consider an instance $\mathfrak{I}_1$ of \textsc{NMWTS}:
given $3m$ numbers $x_1, x_2, \ldots, x_m$, $y_1, y_2, \ldots, y_m$ and $z_1, z_2, \ldots, z_m$,
does there exist two permutations $\sigma_1$ and $\sigma_2$ of $\{1, 2, \ldots, m \}$,
such that $x_i + y_{\sigma_1(i)} = z_{\sigma_2(i)}$ for $1 \leq i \leq m$?
Because \textsc{NMWTS} is NP-complete in the strong sense, we can encode the $3m$ numbers in unary
and assume that the size of $\mathfrak{I}_1$ is $O(m + M)$, where $M = \max_i \{x_i, y_i, z_i\}$.
We also assume that $\sum_{i=1}^{m} x_i + \sum_{i=1}^{m} y_i = \sum_{i=1}^{m} z_i$,
otherwise $\mathfrak{I}_1$ cannot have a solution.

We build the following instance $\mathfrak{I}_2$ of \HETPARTDEC (we use the formulation
in terms of task weights and processor speeds which is more intuitive):
\begin{itemize}
  \item We define $\n = (M+3) m$ tasks, whose weights are outlined below:\\

  $$\begin{array}{ccccccccccccccccccccccccccc}
  A_1 & \underbrace{111...1} & C & D & | & A_2 & \underbrace{111...1} & C & D & |
  & \ldots  & | & A_m & \underbrace{111...1} & C & D\\
  & M & & & &  & M  & & & & & &  & M\\
  \end{array}
  $$

  Here, $B = 2M$, $C = 5M$, $D = 7M$, and $A_i = B + x_i$ for $1 \leq i \leq m$. To define
  the $a_i$ formally for $1 \leq i \leq n$, let $N = M+3$. We have for $1 \leq i \leq m$:
  $$\left\{
  \begin{array}{l}
  a_{(i-1)N+1} = A_i = B + x_{i}\\
  a_{(i-1)N+j} = 1 \text{~for~} 2 \leq j \leq M+1\\
  a_{iN-1} = C\\
  a_{iN} = D
  \end{array}
  \right.$$

  \item For the number of  processors (and intervals), we choose $\p = 3m$. As for the speeds,
  we let $\speed_i$ be the speed of processor $P_i$ where, for $1 \leq i \leq m$:
  $$\left\{
  \begin{array}{l}
  \speed_{i} = B + z_{i}\\
  \speed_{m +i} = C + M - y_i\\
  \speed_{2m+i} = D\\
  \end{array}
  \right.$$

\end{itemize}
Finally, we ask whether there exists a solution matching the bound $K=1$. Clearly, the size of
$\mathfrak{I}_2$ is polynomial in the size of $\mathfrak{I}_1$. We now show that instance
$\mathfrak{I}_1$ has a solution if and only if instance $\mathfrak{I}_2$ does.

%
%\begin{figure}
%  \centering
%  \includegraphics[height=5.5cm]{fig/proof-nphetero.fig}
%  \caption{The platform used in the reduction for Theorem~\ref{th.nphetero}.}
%  \label{fig.proof}
%\end{figure}

\bigskip

Suppose first that $\mathfrak{I}_1$ has a solution, with permutations $\sigma_1$ and $\sigma_2$
such that $x_i + y_{\sigma_1(i)} = z_{\sigma_2(i)}$. For $1 \leq i \leq m$:
\begin{itemize}
  \item  We map each task $A_i$ and the following $y_{\sigma_1(i)}$
tasks of weight $1$ onto processor $P_{\sigma_2(i)}$.
  \item We map the following $M-y_{\sigma_1(i)}$ tasks of weight $1$ and the next task, of weight $C$,
  onto processor $P_{m+\sigma_1(i)}$.
  \item We map the next task, of weight $D$, onto the processor $P_{2m+i}$.
\end{itemize}
We do have a valid partition of all the tasks into $\p =3m$ intervals. For $1 \leq i \leq m$,
the load and speed of the processors are indeed equal:
\begin{itemize}
  \item  The load of $P_{\sigma_2(i)}$ is $A_i + y_{\sigma_1(i)} = B + x_i + y_{\sigma_1(i)}$
  and its speed  is $B + z_{\sigma_2(i)}$.
  \item  The load of $P_{m+\sigma_1(i)}$ is $M -  y_{\sigma_1(i)} + C$, which is equal to its speed.
  \item The load and speed of $P_{2m+i}$ are both $D$.
\end{itemize}
The mapping does achieve the bound $K=1$, hence a solution to
$\mathfrak{I}_1$.

\medskip
Suppose now that $\mathfrak{I}_2$ has a solution, i.e. a mapping matching the bound $K=1$.
We first observe that $\speed_i < \speed_{m+j} < \speed_{2m+k} = D$ for $1 \leq i,j,k \leq m$.
Indeed $\speed_i = B + z_i \leq B+M = 3M$, $5M \leq \speed_{m+j} = C+M-y_j \leq 6M$ and $D=7M$.
Hence each of the $m$ tasks of weight $D$ must be assigned to a processor of speed $D$, and
it is the only task assigned to this processor. These $m$ singleton assignments divide
the set of tasks into $m$ intervals, namely the set of tasks before the first task
of weight $D$, and the $m-1$ sets of tasks lying between two consecutive
tasks of weight $D$. The total weight of each of these $m$ intervals
is $A_i + M + C > B + M + C = 10M$, while the largest speed of the $2m$ remaining processors
is $6M$.  Therefore
each of them must be assigned to at least $2$ processors each. However, there remains only
$2m$ available processors, hence each interval is assigned exactly $2$ processors.

Consider such an interval $A_i~111...1~C$ with $M$ tasks of weight $1$, and let $P_{i_1}$
and $P_{i_2}$
be the two processors assigned to this interval. Tasks $A_i$ and $C$ are not assigned
to the same processor (otherwise the whole interval would). So $P_{i_1}$ receives task $A_i$ and
$h_i$ tasks of weight $1$ while $P_{i_2}$ receives $M-h_i$ tasks of weight $1$ and task $C$.
The weight of $P_{i_2}$ is $M - h + C \geq C = 5M$ while
$\speed_i \leq 3M$ for $1 \leq i \leq m$. Hence $P_{i_1}$ must be some $P_i$, $1 \leq i \leq m$
while $P_{i_2}$ must be some $P_{m+j}$, $1 \leq j \leq m$. Because this holds true on each interval,
this defines two
permutations $\sigma_2(i)$ and $\sigma_1(i)$ such that $P_{i_1} = P_{\sigma_2(i)}$ and
$P_{i_2} = P_{\sigma_1(i)}$. Because the bound $K=1$ is achieved, we
have:\\
$\bullet \quad A_i + h_i = B + x_i + h_i \leq B + z_{\sigma_2(i)}$\\
$\bullet \quad M - h_i + C \leq C + M - y_{\sigma_1(i)}$

Therefore $y_{\sigma_1(i)} \leq h_i$ and $x_i + h_i \leq
z_{\sigma_2(i)}$, and
$$\sum_{i=1}^{m} x_i + \sum_{i=1}^{m} y_i \leq \sum_{i=1}^{m} x_i + \sum_{i=1}^{m} h_i \leq \sum_{i=1}^{m} z_i$$

By hypothesis, $\sum_{i=1}^{m} x_i + \sum_{i=1}^{m} y_i = \sum_{i=1}^{m} z_i$, hence
all the previous inequalities are tight, and in particular $\sum_{i=1}^{m} x_i +
\sum_{i=1}^{m} h_i = \sum_{i=1}^{m} z_i$.

We can deduce that  $\sum_{i=1}^{m} y_i = \sum_{i=1}^{m} h_i =
\sum_{i=1}^{m} z_i - \sum_{i=1}^{m} x_i$, and since
$y_{\sigma_1(i)} \leq h_i$ for all~$i$, we have $y_{\sigma_1(i)} =
h_i$ for all~$i$.

Similarly, we deduce that $x_i + h_i = z_{\sigma_2(i)}$ for all $i$,
and therefore  $x_i + y_{\sigma_1(i)} = z_{\sigma_2(i)}$.

Altogether, we have found a solution for $\mathfrak{I}_1$, which concludes the proof.

%Please refer to the Appendix.
\end{proof}

This important result leads to the NP-completeness of the period
minimization problem.

\begin{theorem}
\label{th.npc2c} %For \COMHOM platforms, the (decision problem associated to the)
%\INTERVAL optimization problem
The period minimization problem for pipeline graphs is NP-complete.
\end{theorem}

\begin{proof} %paragraph{Back to the proof of Theorem~\ref{th.npc2c}.}
Obviously, the optimization
problem belongs to the class NP.
Any instance of the \HETPART problem with $\n$ tasks $a_i$, $p$
processor speeds $\speed_i$ and bound $K$ can be converted into an
instance of the mapping problem with $\n$ stages of weight $\calc_i =
a_i$, letting all communication costs $\comm_i = 0$,
targeting a \COMHOM platform with the same $p$ processors and
homogeneous links of bandwidth $\bw=1$,
and trying to achieve a period
not greater than $K$. This concludes the proof. %\hfill $\square$
\end{proof}

Since the period minimization problem is NP-hard, all bi-criteria
problems are NP-hard.

\section{Heuristics}
\label{sec.heuristics}

%In this section several heuristics are for \COMHOM platforms are presented. We restrict to
%such platforms because, as already pointed out in Section~\ref{sec.intro}, clusters made
%of different-speed processors interconnected by either plain Ethernet or a high-speed
%switch constitute the typical experimental platforms in most academic or industry
%research departments.

The bi-criteria optimization problem is NP-hard,
this is why we propose in this section several polynomial
heuristics to tackle the problem.
In the following, we denote by $\n$ the number of
stages, and by $\p$ the number of processors.

\subsection{Minimizing latency for a fixed period}

In the first set of heuristics, the period is fixed a priori,
and we aim at minimizing the latency
while respecting the prescribed period.
All the following heuristics sort processors by non-increasing speed, and
start by assigning all the stages to the first (fastest)
processor in the list. This processor becomes \emph{used}.

\begin{description}
\item[H1-Sp mono P: Splitting mono-criterion --]% This heuristic sorts the
%processors by decreasing speed, and starts by assigning all the stages to the first
%processor in the list. This processor becomes used. Then, at
At each step, we select the
used processor~$j$ with the largest period and we try to split its stage interval,
giving some stages to the next fastest processor $j'$ in the list (not yet used). This
can be done by splitting the interval at any place, and either placing the first part of
the interval on $j$ and the remainder on $j'$, or the other way round. The solution which
minimizes $max(period(j),period(j'))$ is chosen if it is better than the original
solution. Splitting is performed as long as we have not reached
the fixed period or until we cannot improve the period anymore.

\item[H2a-3-Explo mono: 3-Exploration mono-criterion --] At each step
we select the used processor~$j$ with the largest period and we split its interval into three parts.
For this purpose we try to map two parts of the interval on the next pair of fastest processors in the list, $j'$ and $j''$, and to keep the third part on processor $j$. Testing all possible permutations and all possible positions where to cut, we choose the solution that minimizes $max(period(j),period(j'),period(j''))$.

\item[H2b-3-Explo bi: 3-Exploration bi-criteria --] In this heuristic
the choice of where to split is more elaborated: it depends not only
of the period improvement, but also of the latency increase.
Using the same splitting mechanism as in {\bf 3-Explo mono}, we select
the solution that minimizes $max_{i\in\{j,j',j''\}}(\frac{\Delta
latency}{\Delta period(i)})$.
Here $\Delta latency$ denotes the difference between the global
latency of the solution before the split and after the split. In the
same manner $\Delta period(i)$ defines the difference between the
period before the split (achieved by processor $j$) and the new period
of processor $i$.

\item[H3-Sp bi P: Splitting bi-criteria --] This
heuristic uses a binary search over the latency. For this purpose at
each iteration we fix an authorized
increase of the optimal latency (which is obtained by mapping all stages on
the fastest processor), and
we test if we get a feasible solution via splitting. The splitting
mechanism itself is quite similar to {\bf H1 Sp mono P} except
that we choose the solution that minimizes $max_{i
\in\{j,j'\}}(\frac{\Delta latency}{\Delta period(j)})$ within the
authorized latency increase to decide where to split.
While we get a feasible solution, we reduce the authorized latency
increase for the next iteration of the binary search, thereby aiming at minimizing the mapping global latency.
\end{description}

\subsection{Minimizing period for a fixed latency}

In this second set of heuristics, latency is fixed, and we try to
achieve a minimum period while respecting the latency constraint. As
in the heuristics described above, first of all we sort processors
according to their speed and map all stages on the fastest processor.
The approach used here is the converse of the heuristics where we fix
the period, as we start with an optimal solution concerning
latency. Indeed, at each step we downgrade the solution with respect
to its latency but improve it regarding its period.

\begin{description}
\item[H4-Sp mono L: Splitting mono-criterion --] This heuristic uses
the same method as  {\bf H1-Sp mono P} with a different break
condition. Here splitting is performed as long as we do not exceed the
fixed latency, still choosing the solution that minimizes
$max(period(j),period(j'))$.

\item[H5-Sp bi L: Splitting bi-criteria --] This variant of the
splitting heuristic works similarly to {\bf H4 Sp mono L}, but at each
step it chooses the solution which minimizes
$max_{i\in\{j,j'\}}(\frac{\Delta latency}{\Delta period(i)})$ while
the fixed latency is not exceeded.
\end{description}

The code for all these heuristics can be found on the Web at:\\
\centerline{\url{http://graal.ens-lyon.fr/~vrehn/code/multicriteria/}}

\section{Experiments}
\label{sec.experiments}

Several experiments have been conducted in order to assess the
performance of the
heuristics described in Section~\ref{sec.heuristics}. First we describe the experimental
setting, then we report the results, and finally we provide a summary.

\subsection{Experimental setting}

We have
generated a set of random
applications with
%$\n=1$ to $50$ stages, and two sets of random
%\COMHOM platforms, one set with
%$\p=10$ processors and the other with $\p=100$ processors.
$\n \in \{5,10,20,40\}$ stages and a set of random \COMHOM platforms
with $\p=10$ or $\p = 100$ processors.

%The first case corresponds to
%a situation in which several stages are likely to be mapped on the
%same processor because there are much fewer processors than
%stages. However, in the second case, we expect the mapping to be a
%\ONETOONE, except when communications are really costly.

%The heuristics have been designed for \COMHOM platforms, so we restrict to such
%platforms in these experiments.
In all the experiments, we fix $\bw=10$ for the link
bandwidths. Moreover, the speed of each processor is randomly chosen
as an integer
between~$1$ and~$20$. We keep the latter range of variation throughout
the experiments,
while we vary the range of the application parameters from one set of
experiments to the
other. Indeed, although there are four categories of parameters to
play with, i.e. the
values of $\comm$, $\calc$, $\speed$ and $\bw$, we can see from
equations~(\ref{eq.interval.period}) and~(\ref{eq.interval.latency})
that only the relative ratios $\frac{\comm}{\bw}$ and
$\frac{\calc}{\speed}$ have an impact on the performance.

Each experimental value reported in the following has been calculated
as an average over $50$ randomly chosen application/platforms
pairs. For each of these pairs, we report the performance of the six
heuristics described in Section~\ref{sec.heuristics}.

We report four main sets of experiments conducted both for $\p = 10$ and $\p = 100$
processors. For each experiment, we vary some key
application/platform parameter to assess the impact of this parameter
on the performance of the heuristics.

The first two experiments deal
with applications where communications and computations have the same
order of magnitude, and we study the impact of the degree of
heterogeneity of the communications, i.e. of the variation range of
the $\comm$ parameter:
\begin{itemize}
  \item {\bf (E1): balanced communication/computation, and
homogeneous communications.} \label{sec.exp1}
In the first set of experiments, the application communications are
homogeneous, we fix $\comm_i=10$ for $i=0..\n$. The computation time
required by each stage is randomly chosen between $1$ and $20$. Thus,
communications and computations are balanced within the application.

  \item {\bf (E2): balanced communications/computations, and
heterogeneous communications.}
\label{sec.exp2} In the second set of experiments, the application communications are heterogeneous,
chosen randomly between $1$ and $100$. Similarly to Experiment~1, the computation time
required by each stage is randomly chosen between $1$ and $20$. Thus, communications
and computations are still relatively balanced within the application.
\end{itemize}

The last two experiments deal with imbalanced applications: the
third experiment assumes large computations (large value of the $\calc$ to $\comm$ ratio), and the fourth one reports results for small computations
(small value of the $\calc$ to $\comm$ ratio):
\begin{itemize}
\item {\bf (E3): large computations.} \label{sec.exp3}
In this experiment, the applications are much more demanding on
computations than on
communications, making communications negligible with respect to computation
requirements. We choose the communication time between $1$ and $20$, while the
computation time of each application is chosen between $10$ and~$1000$.
\item {\bf (E4): small computations.} \label{sec.exp4}
The last experiment is the opposite to Experiment 3 since
computations are now negligible compared to communications. The
communication time is still chosen between $1$ and $20$, but the
computation time is now chosen between $0.01$ and~$10$.
\end{itemize}

\subsection{Results}

Results for the entire set of experiments can be found on the Web at
\url{http://graal.ens-lyon.fr/~vrehn/code/multicriteria/}. In the
following we only present the most significant plots.

\subsubsection{With $\p = 10$ processors}

\begin{figure}
   \centering
   \subfigure[10 stages.]{
     \includegraphics[angle=270,width=0.46\textwidth]{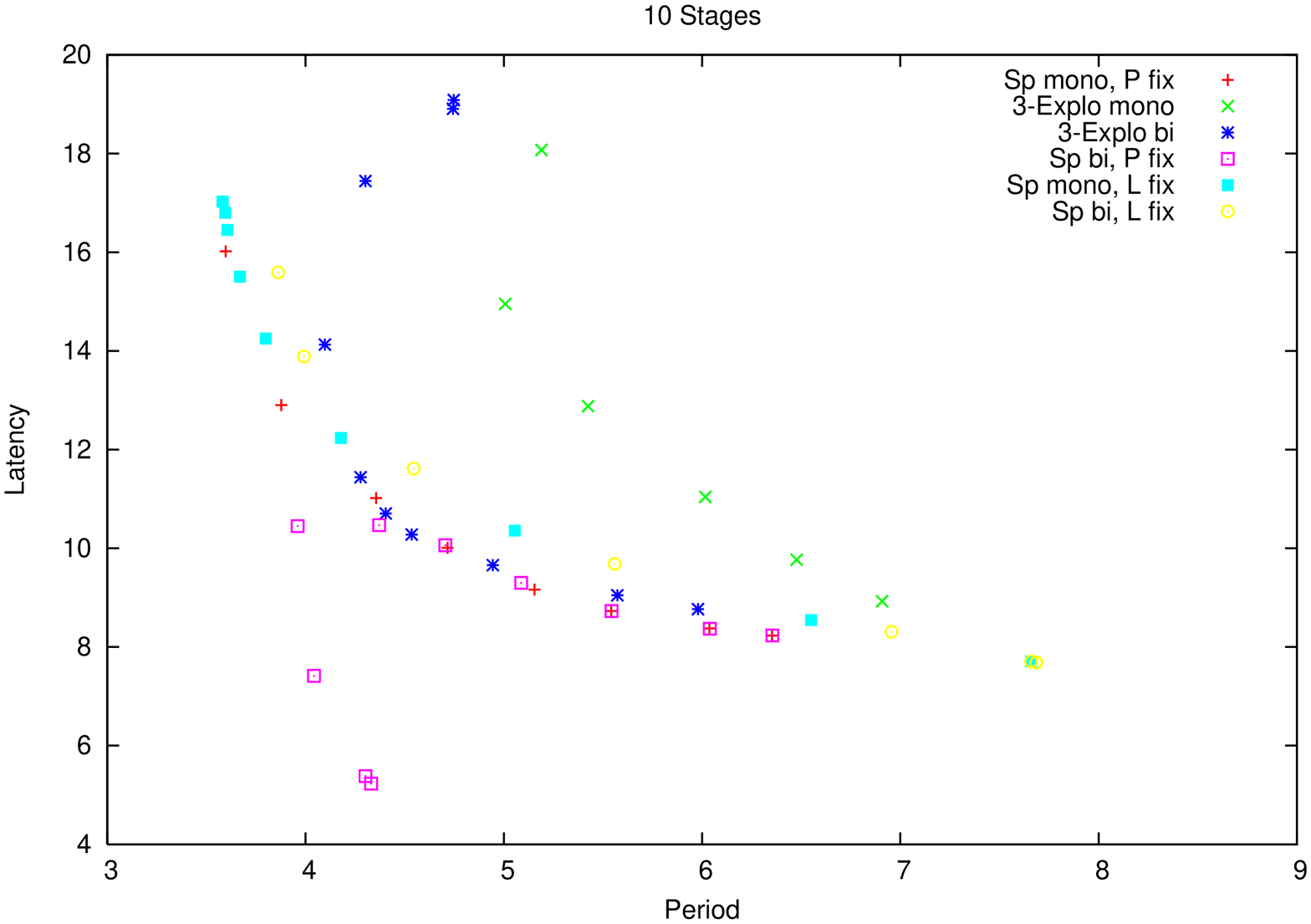}
     \label{fig:ex1-10}
   }$\quad$
   \subfigure[40 stages.]{
     \includegraphics[angle=270,width=0.46\textwidth]{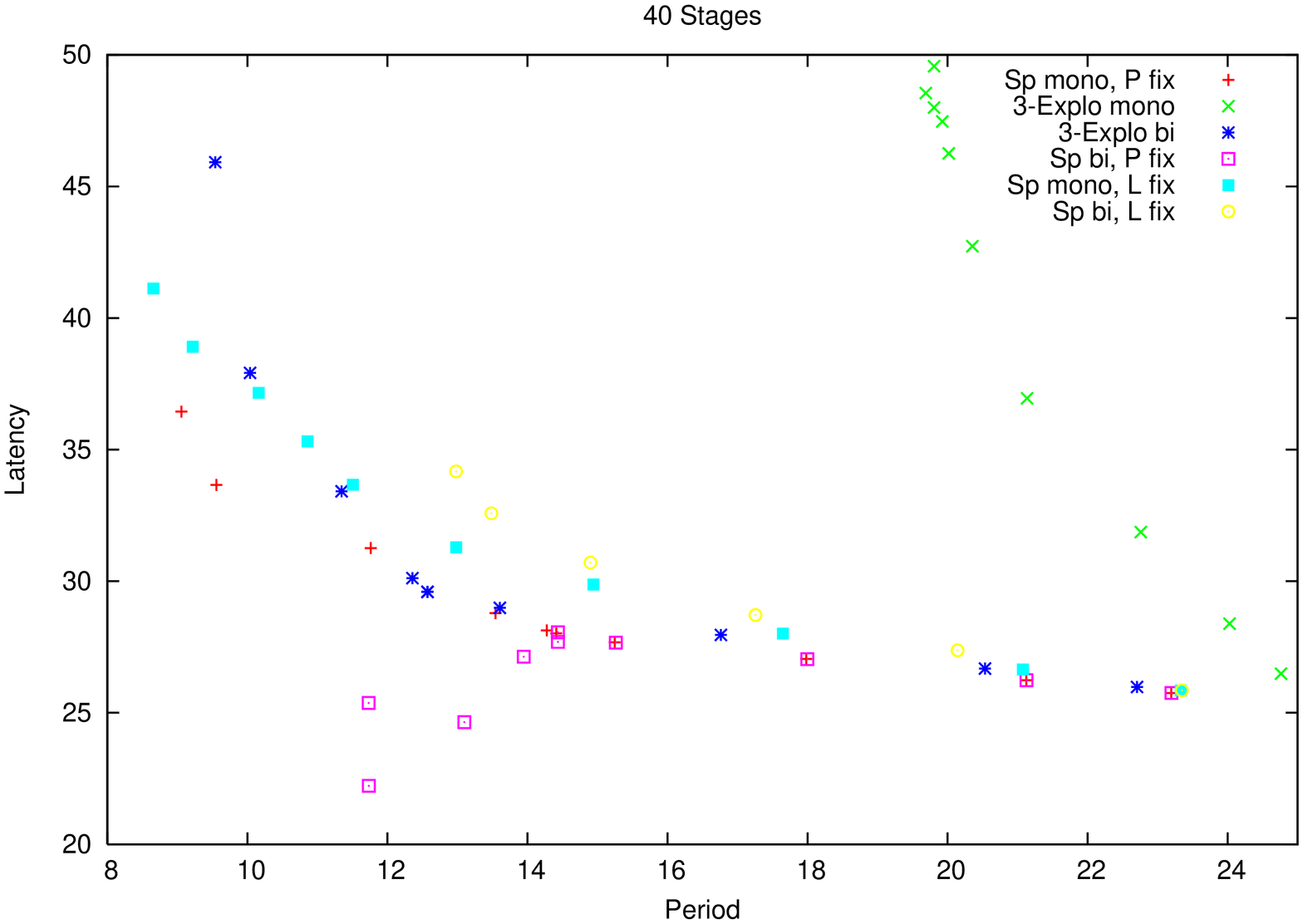}
     \label{fig:ex1-40}
   }
   \caption{(E1) Balanced communications/computations, and
homogeneous communications.}
\label{fig:ex1}
\end{figure}

For (E1) we see that all heuristics follow the same curve shape,
with the exception of heuristic {\bf Sp bi P}
(cf. Figure~\ref{fig:ex1}), which has a different behavior. We observe
this general behavior of the different heuristics in all the
experiments.  The heuristic {\bf Sp bi P} initially finds a solution
with relatively small period and latency, and then tends to
increase both. The other five heuristics achieve small period times at
the price of long latencies and then seem to converge to a
somewhat shorter latency.
We notice that the simplest splitting heuristics perform very well:
{\bf Sp mono P} and {\bf Sp~mono~L} achieve the best period, and {\bf
Sp mono P} has the lower latency. {\bf Sp~bi~P} minimizes the latency
with competitive period sizes. Its counterpart {\bf Sp bi L} performs
poorly in comparison. {\bf 3-Explo mono} and {\bf Sp bi L} cannot keep
up with the other heuristics (but the latter achieves better results
than the former). In the middle range of period values, {\bf 3-Explo
bi} achieves comparable latency values with those of {\bf Sp mono P}
and {\bf Sp bi P}.

\begin{figure}[t]
   \centering
   \subfigure[10 stages.]{
     \includegraphics[angle=270,width=0.46\textwidth]{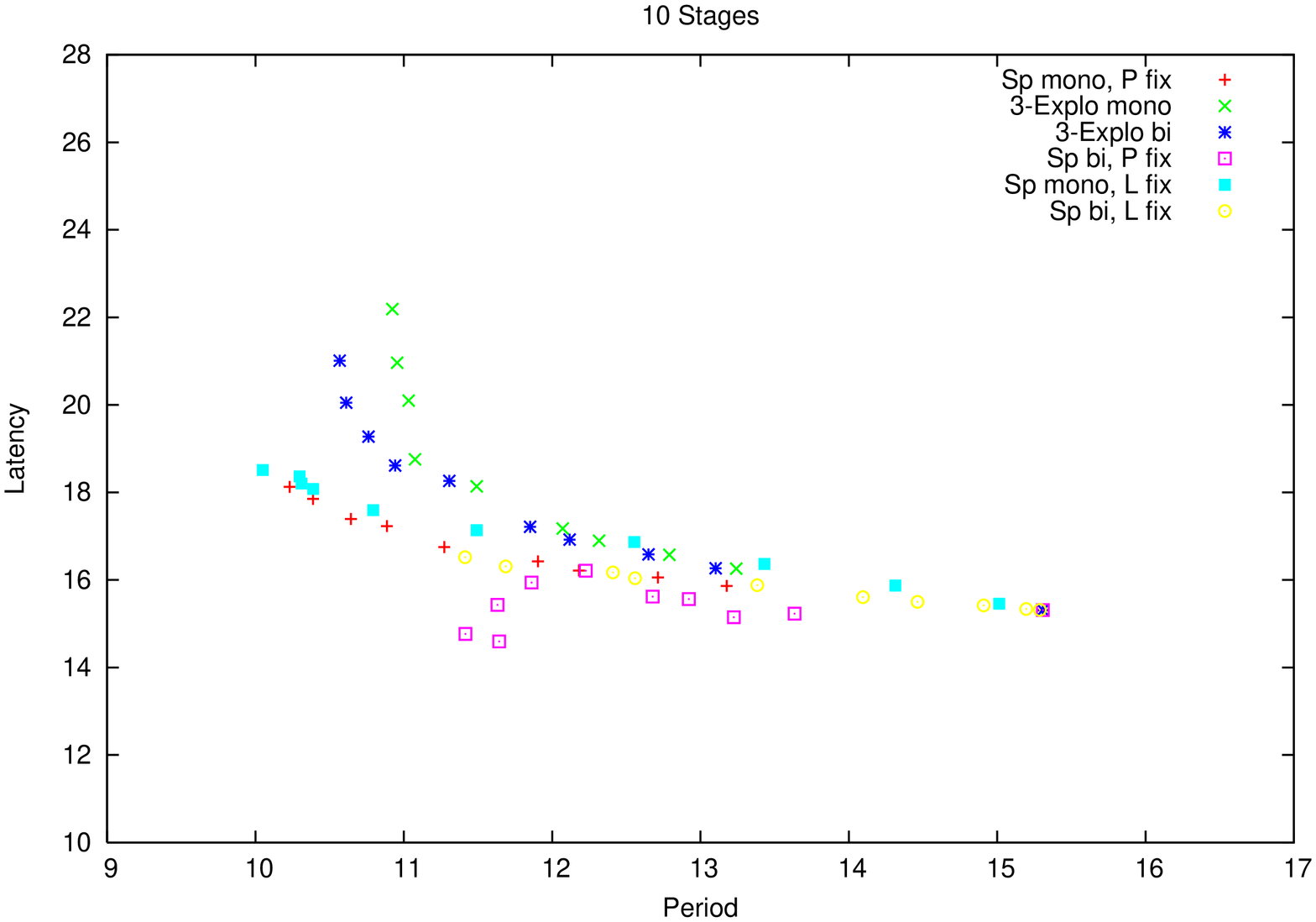}
     \label{fig:ex2-10}
   }$\quad$
   \subfigure[40 stages.]{
     \includegraphics[angle=270,width=0.46\textwidth]{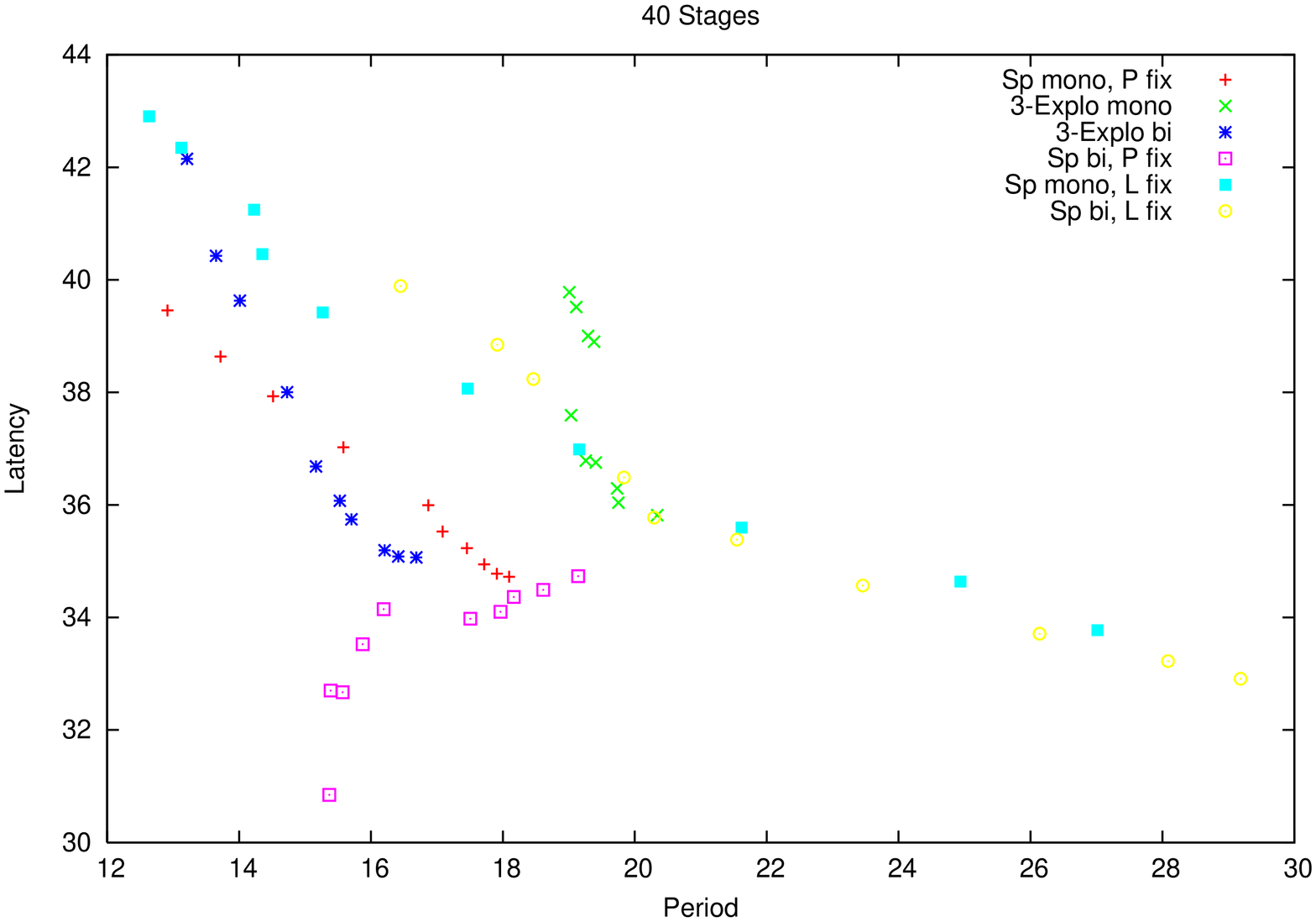}
     \label{fig:ex2-20}
   }
   \caption{(E2) Balanced communications/computations, and
heterogeneous communications.}
\end{figure}

For (E2), if we leave aside {\bf Sp bi P}, we see that {\bf Sp mono P}
outperforms the other heuristics almost everywhere with the following
exception: with $40$ stages and a large fixed period, {\bf
3-Explo bi} obtains the better results. {\bf Sp bi P} achieves by far
the best latency times, but the period times are not as good as those
of {\bf Sp bi P} and {\bf 3-Explo bi}. We observe that the
competitiveness of {\bf 3-Explo bi} increases with the
increase of the number of stages. {\bf Sp mono L} achieves period
values just as small  as {\bf Sp mono P} but the corresponding latency
is higher and once again it performs better than its bi-criteria
counterpart {\bf Sp bi L}. The poorest results are obtained by {\bf
3-Explo mono}.

\begin{figure}[tb]
   \centering
   \subfigure[5 stages.]{
     \includegraphics[angle=270,width=0.46\textwidth]{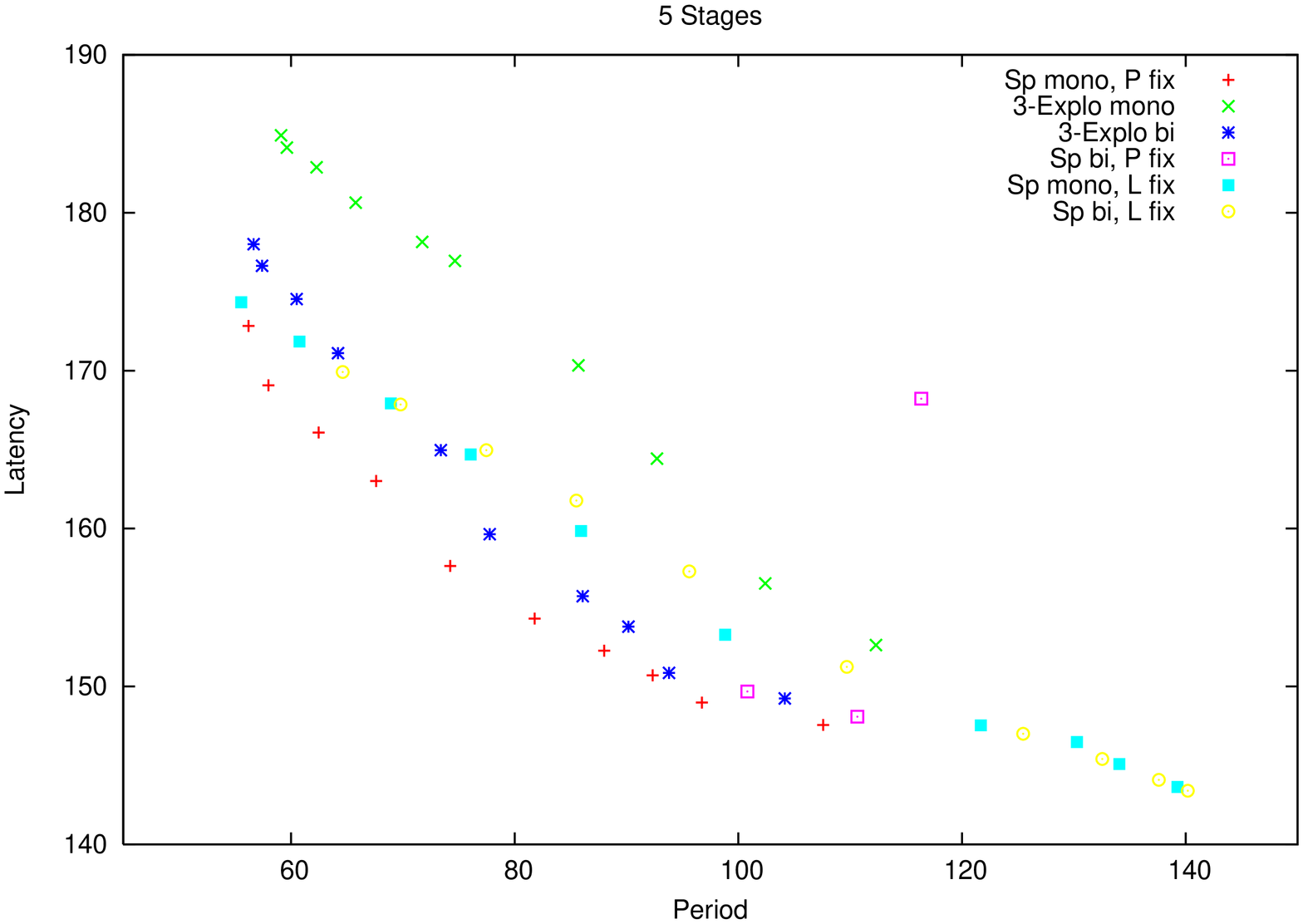}
     \label{fig:ex3-5}
   }$\quad$
   \subfigure[20 stages.]{
     \includegraphics[angle=270,width=0.46\textwidth]{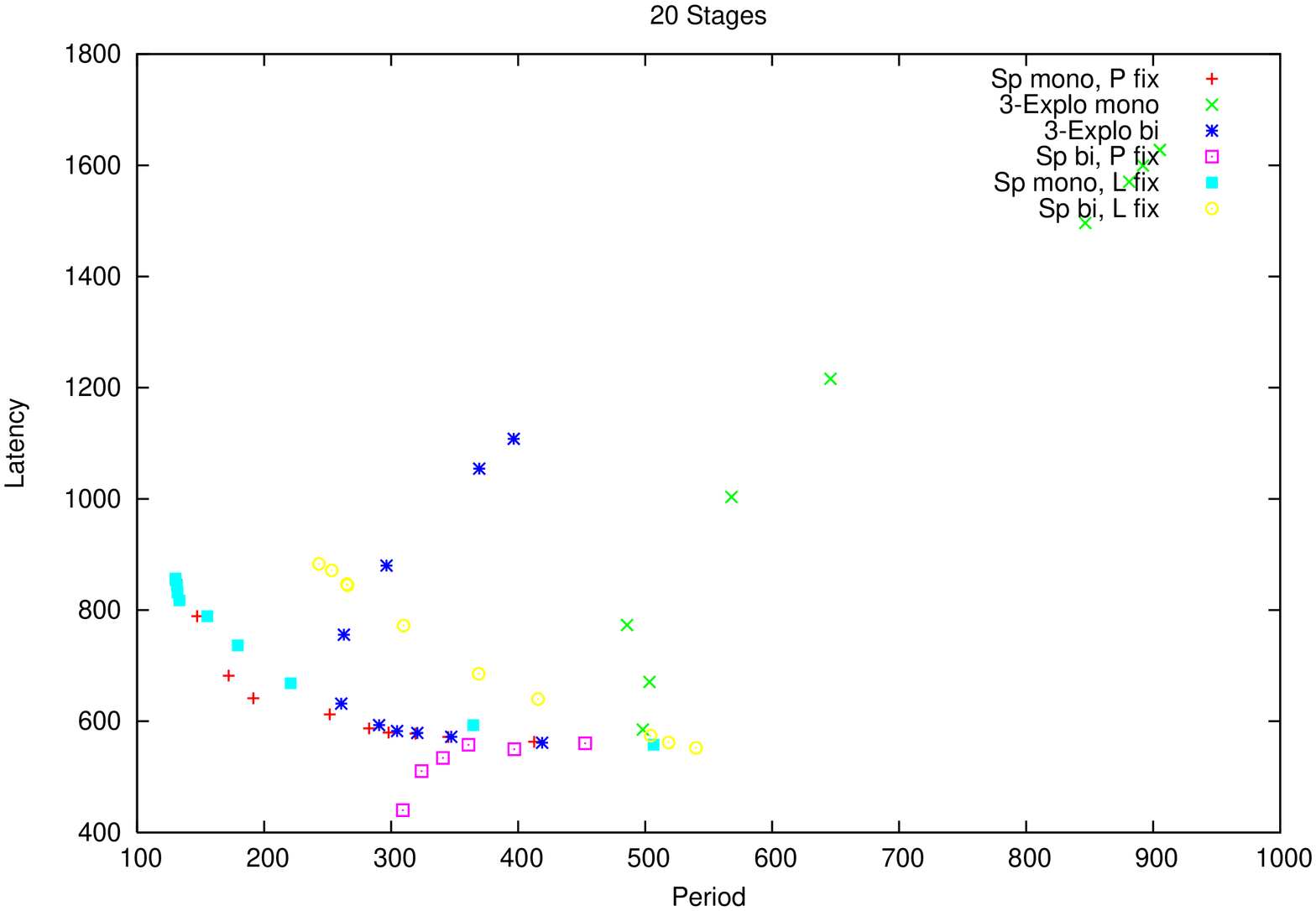}
     \label{fig:ex3-20}
   }
   \caption{(E3) Large computations.}
\end{figure}

The results of (E3) are much more
scattered than in the other experiments (E1, E2 and E4) and this
difference even increases with rising $\n$. When $\n=5$, the results
of the different heuristics are almost parallel so that we can state the
following hierarchy: {\bf Sp mono P}, {\bf 3-Explo bi}, {\bf Sp mono
L}, {\bf Sp bi L} and finally {\bf 3-Explo mono}. For this experiment
{\bf Sp bi P} achieves rather poor results.
With the increase of the number of stages $\n$, the performance of
{\bf Sp bi P} gets better
and this heuristic achieves the best latency, but its period values
cannot compete with {\bf Sp mono P} and {\bf 3-Explo bi}. These latter
heuristics achieve very good results concerning period durations. On
the contrary, {\bf 3-Explo mono} bursts its period and latency
times. {\bf 3-Explo bi} loses its second position for small period
times compared to {\bf Sp mono L}, but when period times are higher it
recovers its position in the hierarchy.

\begin{figure}
   \centering
   \subfigure[5 stages.]{
     \includegraphics[angle=270,width=0.46\textwidth]{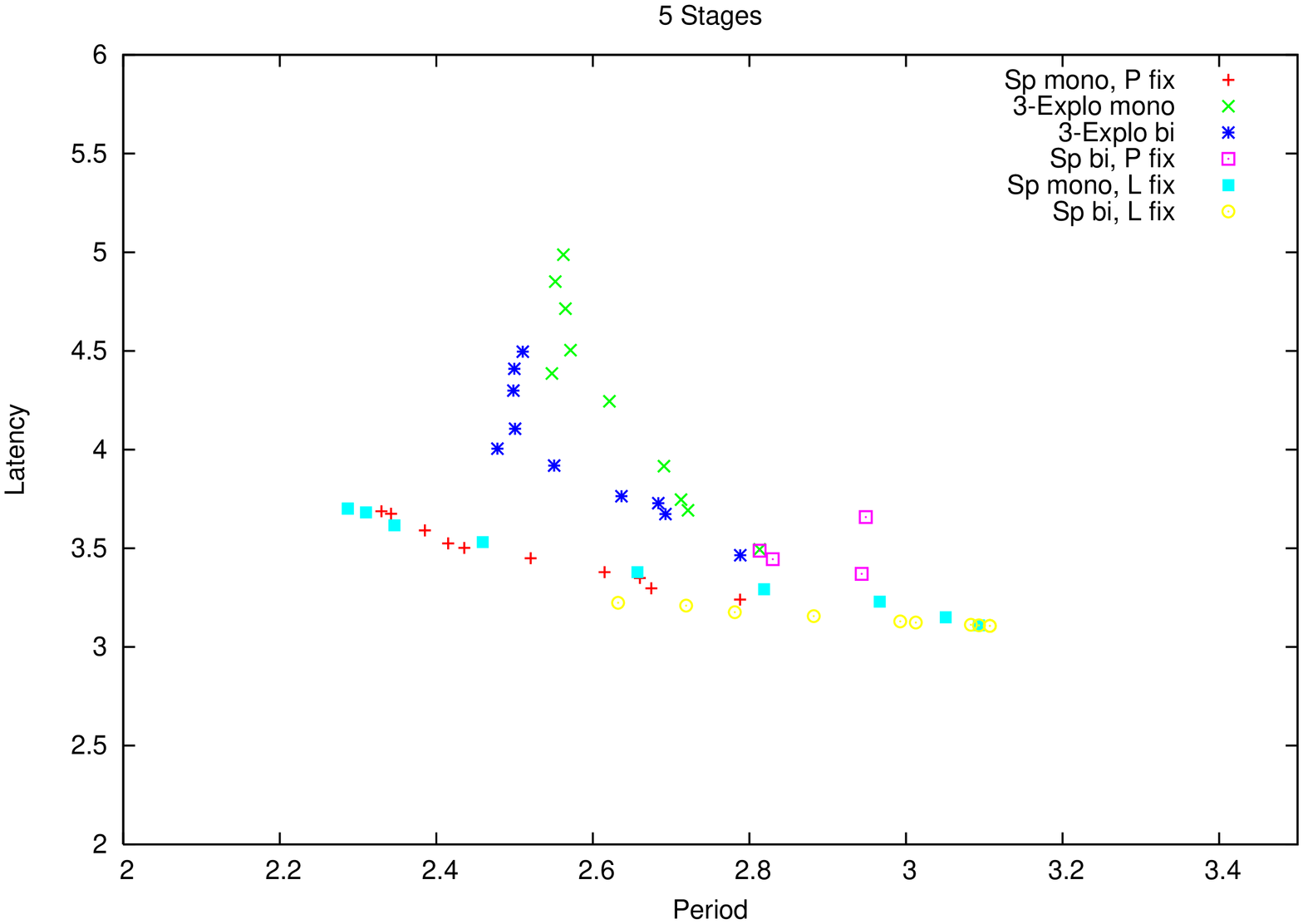}
     \label{fig:ex4-5}
   }$\quad$
   \subfigure[20 stages.]{
     \includegraphics[angle=270,width=0.46\textwidth]{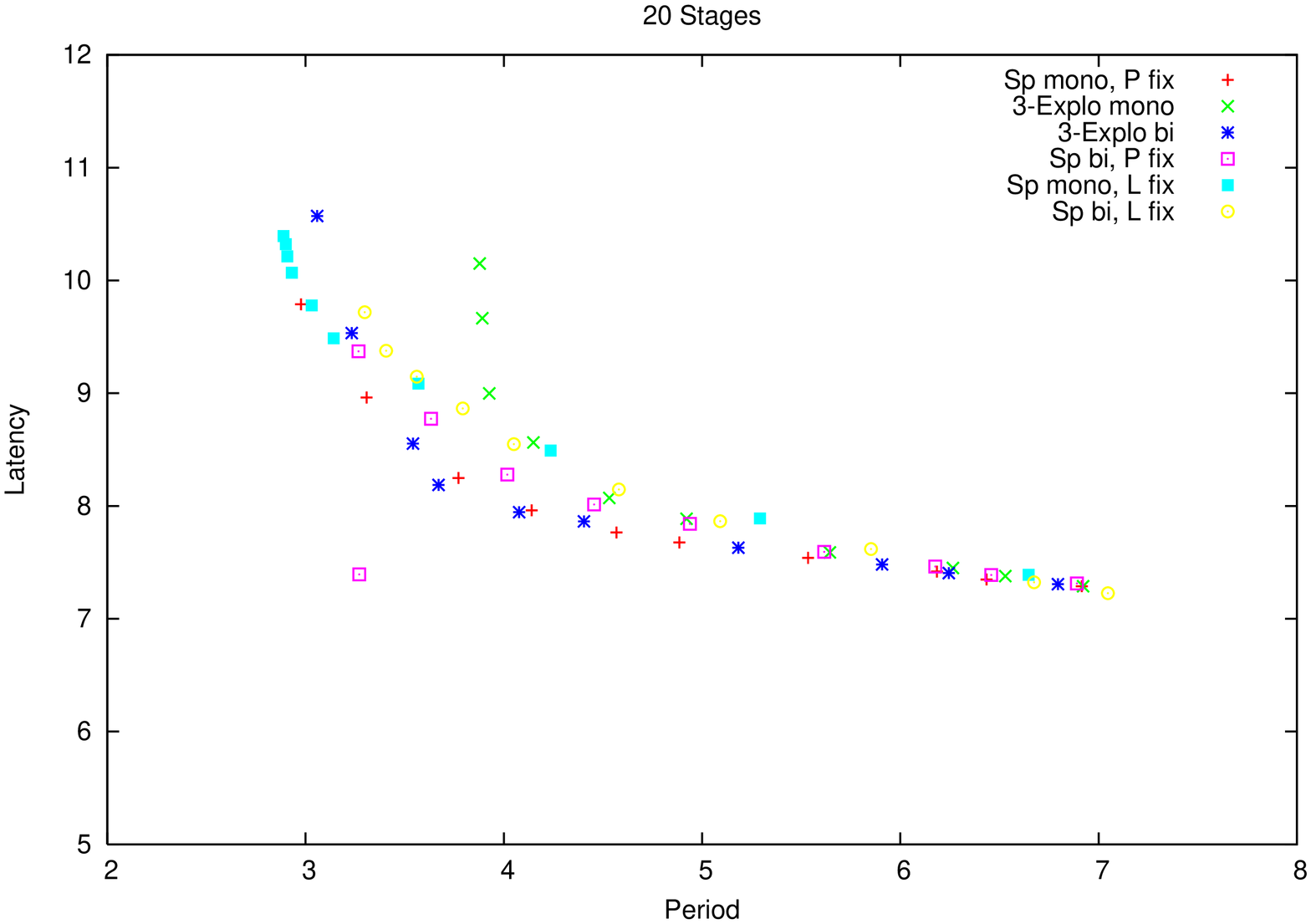}
     \label{fig:ex4-20}
   }
   \caption{(E4) Small computations.}
\end{figure}

In (E4), {\bf 3-Explo mono} performs the
poorest. Nevertheless the gap is smaller than in (E3)
and for high period times and $\n\geq20$, its latency
%%VR: au lieu de n>10 je dirais plutot n \geq 20 vu qu'après 10 les
%% prochains tests sont pour n=20, est-ce correct?
%% ça doit parler de la fig n=20 il me semble...
is comparable to those of the other heuristics. For $\n\geq 20$, {\bf
3-Explo bi} achieves for the first time the best results and the
latency of {\bf Sp bi P} is only one time lower. When $n=5$,
{\bf Sp bi L} achieves the best latency, but the period values are not
competitive with {\bf Sp mono P} and {\bf Sp mono L}, which obtain the
smallest periods (for slightly higher latency times).

In Table~\ref{tab:failure} the {\bf failure thresholds} of the different heuristics are shown. We denote by failure threshold the largest value of the fixed period or latency for which the heuristic was not able to find a solution.
We state that {\bf Sp mono P} has the smallest failure thresholds whereas {\bf 3-Explo mono} has the highest values.
Surprisingly the failure thresholds (for fixed latencies) of the heuristics {\bf Sp mono L} and {\bf Sp bi L} are the same, but their performance differs enormously as stated in the different experiments.

\begin{table}
\centering
\footnotesize{
\begin{tabular}{|r|r| c c c c||r|r| c c c c |}
\hline
Exp.&Heur. & \multicolumn{4}{c||}{Number of stages} & Exp.& Heur.& \multicolumn{4}{c|}{Number of stages}\\
 & & 5&	10&	20&	40& & & 5 & 10 & 20 & 40\\
\hline
E1&H1&	3.0&	3.3&	5.0&	5.0 & E2& H1&	9.7&	10.0&	11.0&	11.0\\
&H2&	3.0&	4.7&	9.0&	18.0 &&H2&	10.3&	10.0&	12.0&	19.0\\
&H3&	3.0&	4.0&	5.0&	5.0&&H3&	10.0&	10.0&	11.0&	11.0\\
&H4&	3.3&	3.3&	6.0&	10.0&&H4&	11.3&	11.0&	13.0&	15.0\\
\hline
&H5&	4.5&	6.0&	13.0&	25.0&&H5&	11.7&	15.0&	22.0&	32.0\\
&H6&	4.5&	6.0&	13.0&	25.0&&H6&	11.7&	15.0&	22.0&	32.0\\
\hline
\hline
E3 &H1&	50.0&	70.0&	100.0&	250.0&E4&H1&	2.2&	2.3&	2.3&	2.3\\
&H2&	50.0&	140.0&	450.0&	950.0&&H2&	2.4&	2.7&	3.7&	7.0\\
&H3&	50.0&	90.0&	250.0&	400.0&&H3&	2.4&	2.7&	3.0&	4.0\\
&H4&	100.0&	140.0&	300.0&	650.0&&H4&	2.8&	2.7&	3.0&	4.0\\
\hline
&H5&	140.0&	270.0&	500.0&	1000.0&&H5&	3.0&	4.0&	7.0&	11.0\\	
&H6&	140.0&	270.0&	500.0&	1000.0&&H6&	3.0&	4.0&	7.0&	11.0\\
\hline
\end{tabular}
}
\caption{Failure thresholds of the different heuristics in the different experiments.}
\label{tab:failure}
\end{table}

\subsubsection{With $\p = 100$ processors}

\begin{figure}
   \centering
   \subfigure[(E1) 40 stages, hom. comms.]{
     \includegraphics[angle=270,width=0.46\textwidth]{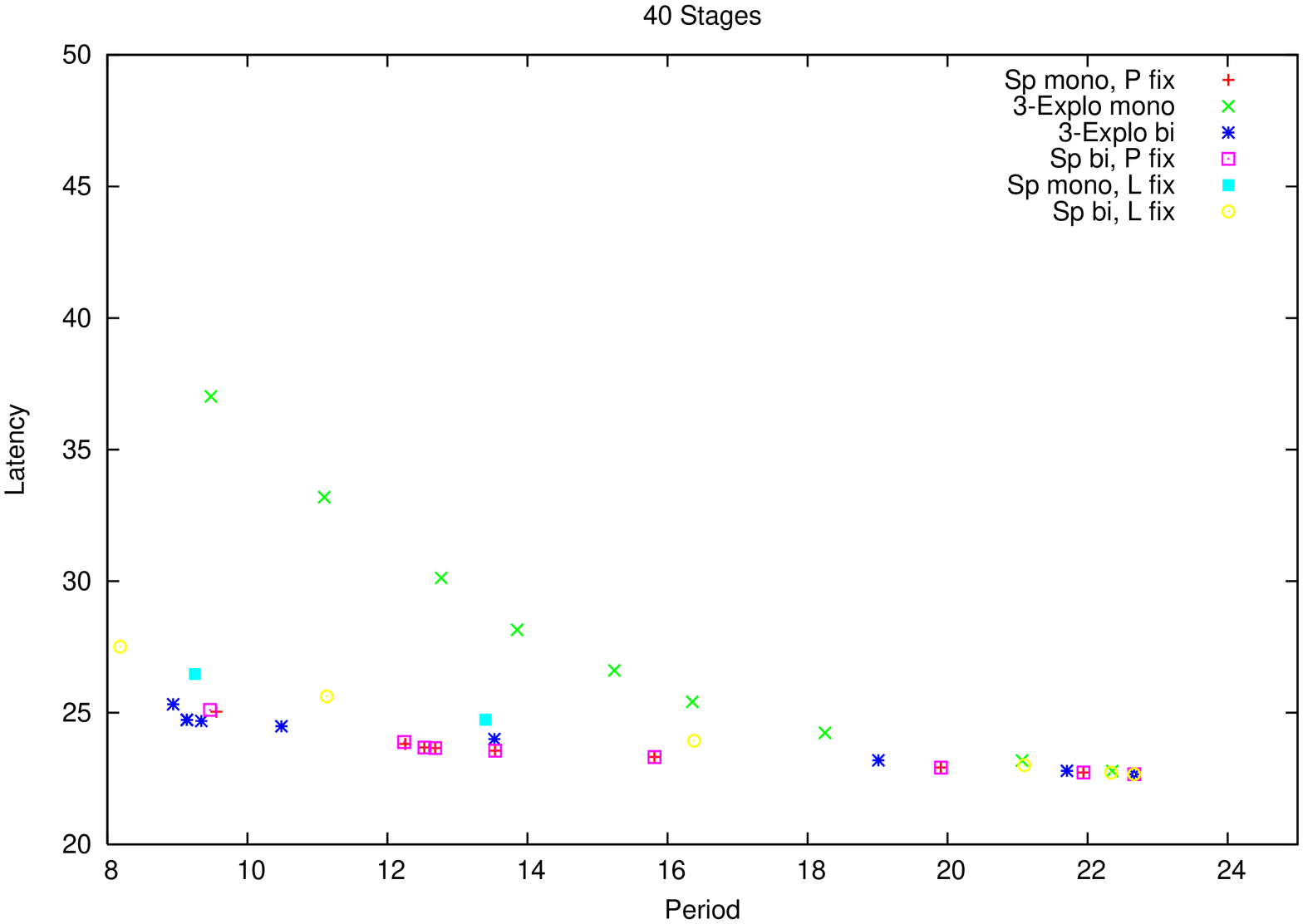}
     \label{fig:E21-40}
   }$\quad$
   \subfigure[(E2) 40 stages, het. comms.]{
     \includegraphics[angle=270,width=0.46\textwidth]{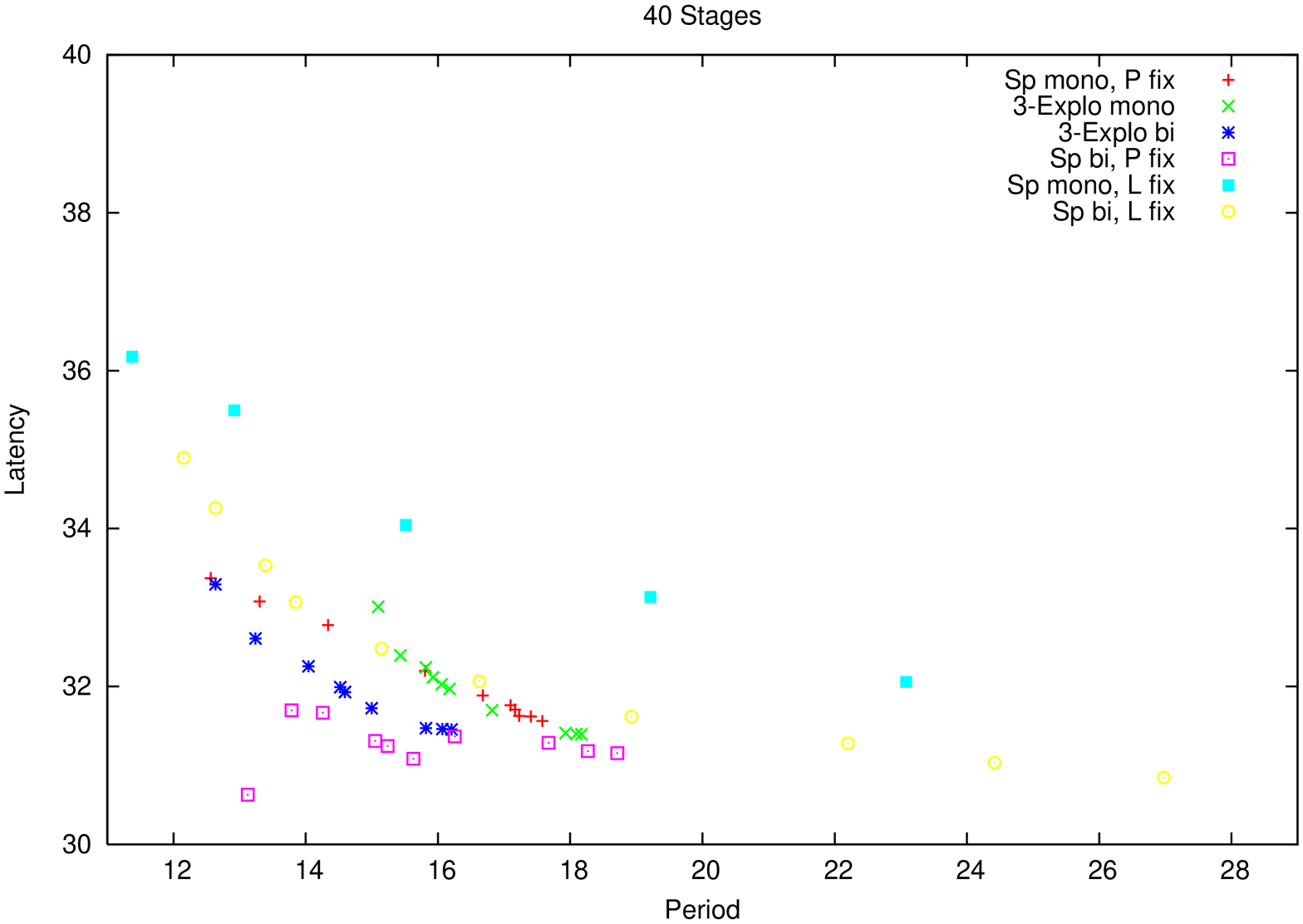}
     \label{fig:E22-40}
   }
   \caption{Extension to 100 processors, balanced communications/computations.}
\end{figure}

Many results are similar with $\p=10$ and $\p=100$ processors, thus we only report
the main differences. First we observe that both periods and latencies are lower with the increasing
 number of processors. This is easy to explain, as all heuristics always choose
fastest processors first, an there is much more choice with $\p=100$.
All heuristics keep their general behavior, i.e. their curve
 characteristics. But the relative performance of some heuristics
 changes dramatically. The results of {\bf 3-Explo mono} are much better, and we do
get adequate latency times (compare
Figures~\ref{fig:ex1-40} and \ref{fig:E21-40}). Furthermore the multi-criteria
heuristics turn out to be much more performant.
An interesting example can be seen in Figure~\ref{fig:E22-40}: all multi-criteria
heuristics outperform their mono-criterion counterparts, even {\bf Sp
bi L}, which never had 
a better performance than {\bf Sp mono L} when $\p=10$.

\begin{figure}
   \centering
   \subfigure[(E3) 10 stages, large computations.]{
     \includegraphics[angle=270,width=0.46\textwidth]{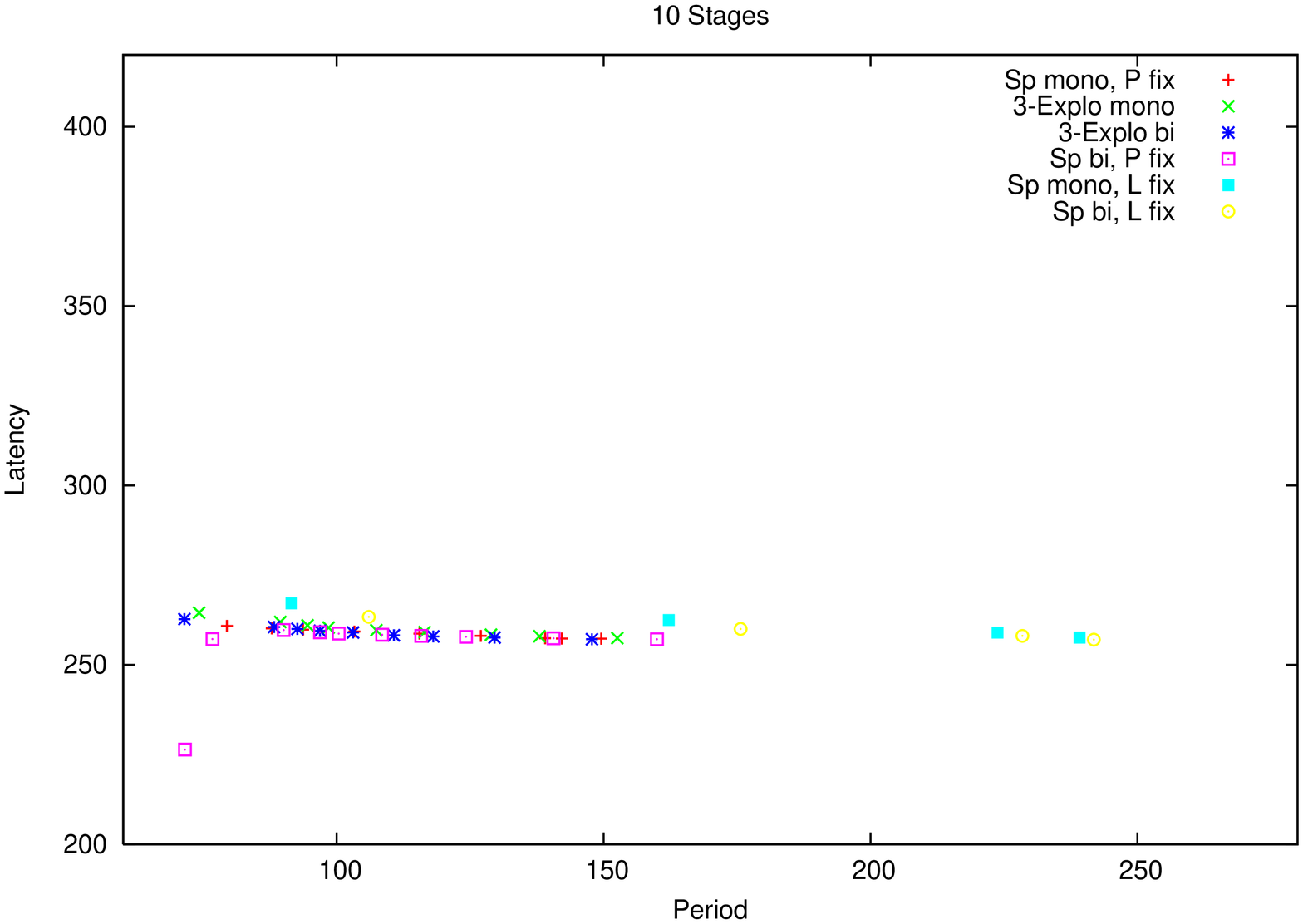}
     \label{fig:E25-10}
   }$\quad$
   \subfigure[(E4) 40 stages, small computations.]{
     \includegraphics[angle=270,width=0.46\textwidth]{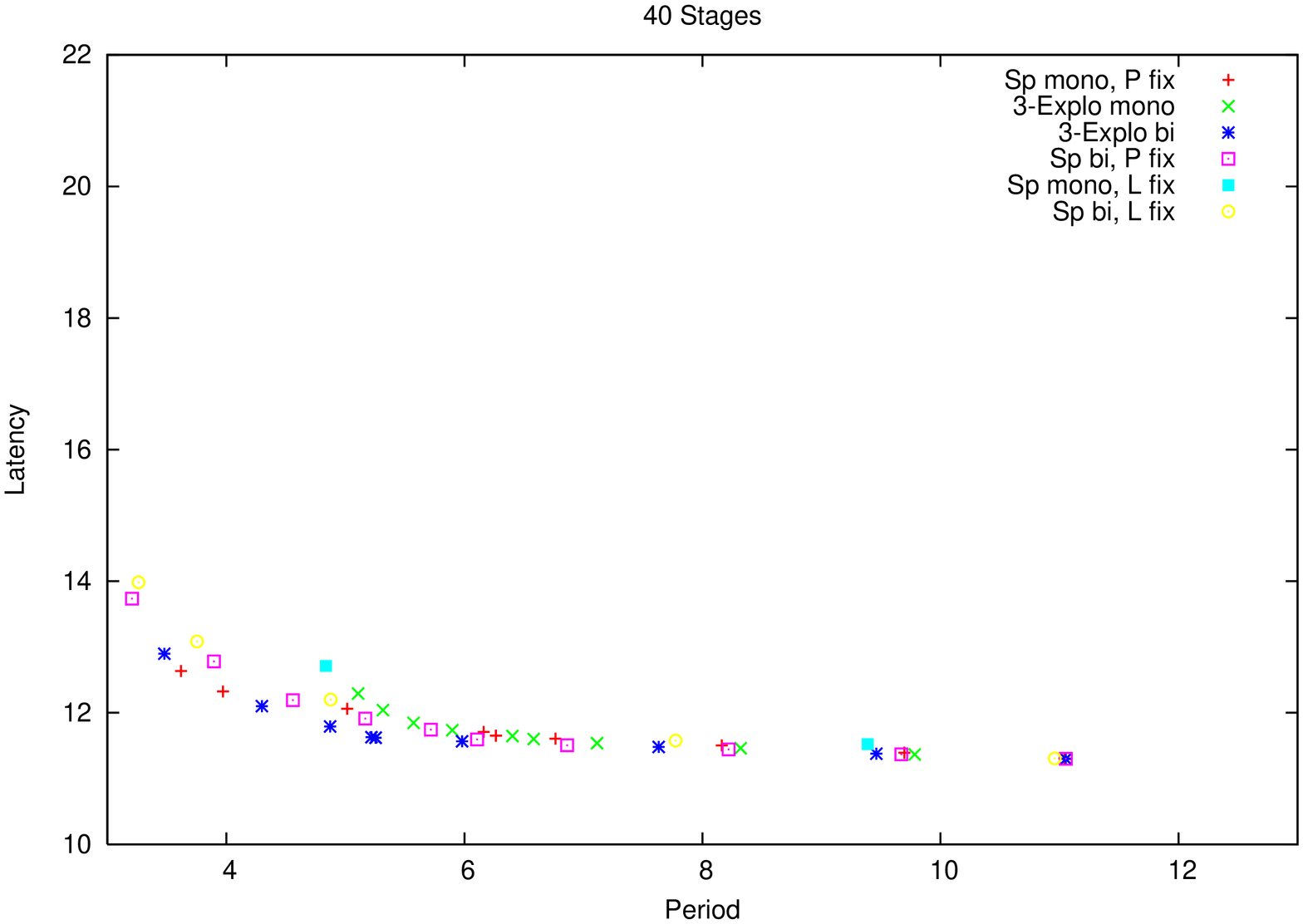}
     \label{fig:E26-40}
   }
   \caption{Extension to 100 processors, imbalanced communications/computations.}
\end{figure}

In the case of imbalanced communications/computations, we observe that
all heuristics achieve almost the same results. The only exception is
the binary-search heuristic {\bf Sp bi P}, which shows a slightly
superior performance as can be seen in Figure~\ref{fig:E25-10}. 
The performance of {\bf 3-Explo bi} depends on the number of
stages. In general it is superseded by {\bf Sp mono P}, when $n\leq
10$, but for $n\geq 20$ {\bf 3-Explo bi} it owns the second position
after {\bf Sp bi L} and even performs best in the configuration small
computations/$n=40$ (see Figure~\ref{fig:E26-40}).

\subsubsection{Summary}

Overall we conclude that the performance of bi-criterion heuristics versus mono-criterion
heuristics highly depends on the number of available processors.

For a small number of processors, the
simple splitting technique which is used
in {\bf Sp mono P} and {\bf Sp mono L} is very competitive 
as it almost always minimizes the period with acceptable latency
values.
The bi-criteria splitting
{\bf Sp bi P} mainly
minimizes latency values at the price of longer periods. Nevertheless
depending upon the application, this heuristics seems to be very
interesting, whenever small latencies are demanded.
On the contrary, its counterpart {\bf Sp bi L} does not provide
convincing results. Finally, both 3-Exploration heuristics do not
achieve the expected performance.

However when increasing he number of available processors, we observe 
a significant improvement
of the behavior of bi-criteria heuristics. {\bf Sp bi L} turns out to outperform the
mono-criterion version and {\bf Sp bi P} upgrades its period times such that it outplays its competitors. 
Finally both 3-Exploration heuristics perform much better and {\bf 3-Explo bi} finds its slot.

\section{Related work}
\label{sec.related}

As already mentioned, this work is an extension of the work of
Subhlok and Vondran~\cite{subhlock-ppopp95,subhlock-spaa96} for
pipeline applications on homogeneous platforms. We extend the
complexity results to heterogeneous platforms.
We have also discussed the relationship with the chains-to-chains
problem~\cite{Bokhari88,Iqbal91,HansenLih92,IqbalBok95,olstad95efficient,PinarAykanat2004}
in Section~\ref{sec.intro}.

Several papers consider the problem of mapping communicating tasks
onto heterogeneous platforms,
but for a different applicative framework. % is different.
In~\cite{TauraChi00}, Taura and Chien consider applications composed
of several copies of the same
task graph, expressed as a DAG (directed acyclic graph). These copies
are to be executed in pipeline fashion.
Taura and Chien also restrict to mapping all instances of
a given task type (which corresponds to a stage in our framework)
onto the same processor. %In other words, they consider the same problem
%as ours, except that the linear pipeline is replaced by a general
%DAG.
Their problem is shown NP-complete,
and they provide an iterative heuristic to determine a good
mapping. At each step, the heuristic refines
the current clustering of the DAG. Beaumont et al.~\cite{c112} consider
the same problem
as Taura and Chien, i.e. with a general DAG, but they allow a given
task type to be mapped onto several processors,
each executing a fraction of the total number of tasks. The problem
remains NP-complete, but becomes
polynomial for special classes of DAGs, such as series-parallel
graphs. For such graphs,
it is possible to determine the optimal mapping owing to an approach
based upon a linear programming
formulation. The drawback with the approach of~\cite{c112} is that the
optimal throughput can only be achieved through
very long periods, so that the simplicity and regularity of the
schedule are lost, while the latency is severely increased.

Another important series of papers comes from the DataCutter
project~\cite{datacutter}.
One goal of this project is to schedule multiple data analysis
operations onto clusters and grids,
decide where to place and/or replicate various
components~\cite{Saltz01,beynon02optimizing,SpencerFB02}.
A typical application is a chain of consecutive filtering operations,
to be executed on a very large data set. The task graphs
targeted by DataCutter are more general than linear pipelines or forks,
but still more regular than arbitrary DAGs, which makes it possible
to design efficient heuristics to solve the previous
placement and replication optimization problems.
However, we point out that a recent paper~\cite{Naga07} targets workflows structured as
arbitrary DAGs and considers bi-criteria optimization problems on homogeneous
platforms. The paper provides many interesting ideas and several
heuristics to solve the general mapping problem.
It would be very interesting to experiment these heuristics on
the simple pipeline mapping problem, and to compare it to our own
heuristics designed specifically for pipeline workflows.

\section{Conclusion}
\label{sec.conclusion}

In this paper, we have studied a difficult bi-criteria mapping problem onto
\COMHOM platforms. We restricted ourselves to the class of
applications which have a pipeline structure, and studied the
complexity of the problem.
To the best of our knowledge, it is the first time that a
multi-criteria pipeline mapping is studied from a theoretical
perspective, while it is quite a standard and widely used pattern in
many real-life applications.

While minimizing the latency is trivial, the problem of minimizing the
pipeline period is NP-hard, and thus the bi-criteria problem is
NP-hard. We provided several efficient polynomial heuristics, either
to minimize the period for a fixed latency, or to minimize the latency
for a fixed period.

These heuristics have been extensively compared through simulation.
%%VR il faudrait ici faire un bref bilan des experiments,
%% j'avais ecrit la conclusion avant les resultats ;-)
Results highly depend on platform parameters such as number of stages
and number of available processors.
Simple mono-criterion splitting heuristics perform very well when there is a limited
number of processors, whereas bi-criterion heuristics perform much better when
 increasing the number of processors. Overall, the introduction of bi-criteria heuristics
was not fully successful for small clusters but turned out to be mandatory to achieve good performance on
larger platforms.

\medskip

There remains much work to extend the results of this paper.
We designed heuristics for \COMHOM platforms, and finding efficient
bi-criteria heuristics was already a challenge. It would be
interesting to deal with fully heterogeneous platforms, but it seems
to be a difficult problem, even for a mono-criterion optimization problem.
In the longer term, we plan to perform real experiments on
heterogeneous platforms, using an already-implemented skeleton
library, in order to compare the effective performance of the
application for a given mapping (obtained with our heuristics) against the
theoretical performance of this mapping.

A natural extension of this work would be to consider other widely used
skeletons. For example, when there is a bottleneck in the pipeline operation
due to a stage which is both computationally-demanding
and not constrained by internal
dependencies, we can nest another skeleton in place of the stage. For
instance a farm or deal skeleton would allow to split the
workload of the initial stage among several processors. Using such deal
skeletons may be either the programmer's decision (explicit nesting
in the application code) or the result of the mapping procedure.
Extending our mapping strategies to automatically identify
opportunities for deal skeletons, and implement these, is
a difficult but very interesting perspective.

%\onehalfspacing
\bibliographystyle{plain}
\bibliography{biblio}

%\newpage
%\section*{Appendix}

%\begin{proof} 
%\end{proof}

\end{document}